\documentclass{vgtc}                          

\ifpdf
  \pdfoutput=1\relax                   
  \usepackage{graphicx}                
  \DeclareGraphicsExtensions{.pdf,.png,.jpg,.jpeg} 
\else
  \usepackage{graphicx}                
  \DeclareGraphicsExtensions{.eps}     
\fi%

\graphicspath{{figures/}{pictures/}{images/}{./}} 

\usepackage{microtype}                 
\PassOptionsToPackage{warn}{textcomp}  
\usepackage{textcomp}                  
\usepackage{mathptmx}                  
\usepackage{times}                     
\usepackage{cite}                      
\usepackage{tabu}                      
\usepackage{booktabs}   

\usepackage[breaklinks=true]{hyperref}
\usepackage{amssymb} 

\onlineid{1092}

\vgtccategory{Technique or Algorithm}

\vgtcinsertpkg

\title{MinMaxLTTB: Leveraging MinMax-Preselection to Scale LTTB} 
    
\newcommand*\samethanks[1][\value{footnote}]{\footnotemark[#1]}

\author{
    Jeroen Van Der Donckt 
    \thanks{contributed equally} 
    \thanks{e-mail: (firstname)(dot)(lastname)(at)ugent(dot)be}
    \and Jonas Van Der Donckt  
    \samethanks[1]
    \and Michael Rademaker
    \and Sofie Van Hoecke
}
\affiliation{\scriptsize IDLab, Ghent University - imec, Belgium}

\teaser{
  \centering
  \includegraphics[width=\linewidth]{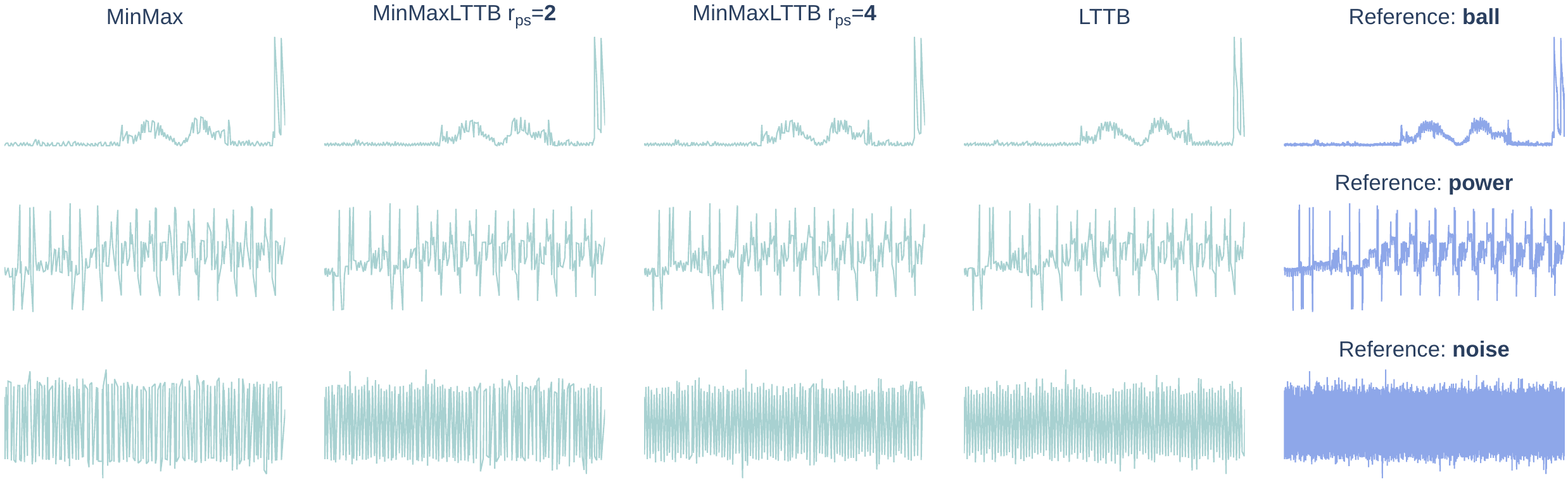}
  \caption{Visual analysis of MinMax (column 1), \texttt{MinMaxLTTB} with preselection ratio $r_{ps} \in \{2, 4\}$ (column 2-3), and LTTB (column 4). Each row corresponds to a unique template, with the rightmost column presenting the reference images, constructed from 200k data points. The four adjacent aggregations utilize an equal number of output samples ($n_{out}=200$). 
  Columns 2 and 3 illustrate the effect of the preselection ratio $r_{ps}$ for MinMaxLTTB. 
  Visualization properties used include: Plotly as toolkit, line-width of 2 pixels, and anti-aliasing enabled. A \href{https://github.com/predict-idlab/MinMaxLTTB/blob/main/gifs/teaser_preselection_ratio.gif}{GIF} further demonstrates the above visualization for varying values for $n_{out}$.
  }
  \label{fig:teaser}
}

\abstract{
Visualization plays an important role in analyzing and exploring time series data. 
To facilitate efficient visualization of large datasets, downsampling has emerged as a well-established approach. This work concentrates on LTTB (Largest-Triangle-Three-Buckets), a widely adopted downsampling algorithm for time series data point selection. 
Specifically, we propose MinMaxLTTB, a two-step algorithm that marks a significant enhancement in the scalability of LTTB.
MinMaxLTTB entails the following two steps: (i) the MinMax algorithm preselects a certain ratio of minimum and maximum data points, followed by (ii) applying the LTTB algorithm on only these preselected data points, effectively reducing LTTB's time complexity. 
The low computational cost of the MinMax algorithm, along with its parallelization capabilities, 
facilitates efficient preselection of data points. Additionally, the competitive performance of MinMax in terms of visual representativeness also makes it an effective reduction method.
Experiments show that MinMaxLTTB outperforms LTTB by more than an order of magnitude in terms of computation time. Furthermore, preselecting a small multiple of the desired output size already provides similar visual representativeness compared to LTTB.
In summary, MinMaxLTTB leverages the computational efficiency of MinMax to scale LTTB, without compromising on LTTB's favored visualization properties.
The accompanying code and experiments of this paper can be found at \url{https://github.com/predict-idlab/MinMaxLTTB}.
} 

\CCScatlist{
  \CCScatTwelve{Time series}{Line charts}{Downsampling algo\-rithms}{MinMax}--
  \CCScatTwelve{LTTB}{Computational efficiency}{Perception}{Preselection}--
  \CCScatTwelve{Evaluation}{}{}{}
}

\begin{document}



\maketitle


\section{Introduction} 
Visualization is widely recognized as a powerful tool for analyzing and exploring time series data, with line charts proving particularly effective for most tasks~\cite{aigner_visualizing_2007}. As the volume of time series data continues to expand, there is an increasing need for efficient visualization methods capable of handling large datasets~\cite{bikakis_big_2018, godfrey2016interactive, rostislav_performance_javascript_vis}. To address this challenge, downsampling has emerged as a well-established technique that involves either aggregating or selecting a representative subset of the time series~\cite{agrawal_challenges_2015, aigner_visualizing_2007, kwon_sampling_2017}. By reducing the number of data points while preserving the overall shape of the time series, downsampling minimizes network latency and improves rendering time, making it a vital component in numerous widely adopted time series databases~\cite{uber_raskin_aggarwal_2018, timescaleblog_paganini_2023}.

This work specifically focuses on the LTTB (Largest-Triangle-Three-Buckets) algorithm~\cite{steinarsson_downsampling_nodate}, which is a value preserving aggregation method as it downsamples by selecting data points from the original time series~\cite{jugel_m4_2014, jugel_vdda_2016}.
LTTB has gained widespread adoption in industry, with companies like Uber incorporating it as a downsampling function in their M3 metrics platform~\cite{uber_raskin_aggarwal_2018} and TimeScaleDB offering LTTB as a server-side hyperfunction~\cite{timescaleblog_paganini_2023}.

Despite its broad use, LTTB exhibits certain computational limitations that restrict its applicability to massive datasets containing billions of data points. Specifically, LTTB involves expensive operations to compute triangular surfaces for each data point and requires a sequential pass over the data, making it unsuitable for parallelization. 
To overcome these computational challenges, we introduce \textit{MinMaxLTTB}, a two-step approach that (i) employs MinMax-preselection as an initial data reduction step, and (ii) applies LTTB on the preselected data points. In particular, this approach leverages the computational efficiency of the MinMax algorithm to make LTTB more scalable.
Utilizing the visual evaluation framework that we proposed in previous work~\cite{van2023data}, we evaluate the visual representativeness of the proposed MinMaxLTTB algorithm for various MinMax-preselection ratios. These findings provide empirical evidence for choosing an appropriate MinMax-preselection ratio. Furthermore, we assess the performance improvement of MinMaxLTTB over LTTB.
Notably, the presented technique is the default downsampling approach in our open-source time series visualization tool plotly-resampler~\cite{van2022plotly}, which has at the time of writing almost 1 million installations.



\section{Related Work}
This section provides an overview of related work in the field, focusing on the scalability of value preserving data aggregation algorithms for time series line chart visualization. 

\subsection{Time Series Visualization}
Visualization is crucial for exploring time series data, with the human eye frequently being advocated as the ultimate data mining instrument~\cite{lin_visualizing_2005}. 
For the majority of time series analysis tasks, simple visualizations, such as line charts, have proven most effective~\cite{aigner_visualizing_2007}. 
Interactive visualization techniques, as emphasized by Shneiderman's visual information-seeking mantra~\cite{shneiderman_eyes_nodate}, are essential for exploring large time series data by providing an overview, enabling zooming and filtering, and allowing access to details-on-demand~\cite{walker_timenotes_2016}.

Most interactive approaches render visualizations client-side, often in web-based environments~\cite{caldarola2017big, rostislav_performance_javascript_vis}. However, this approach presents two significant challenges when handling large data volumes: (i) considerable network latency due to the transmission of large data volumes, and (ii) poor client-side rendering performance~\cite{agrawal_challenges_2015}. Both aspects limit responsiveness and interactivity, which, as highlighted above, is crucial for effectively exploring time series data. To overcome these limitations, data aggregation has proven to be an effective approach~\cite{kumar_mr_2021, bikakis_big_2018, kwon_sampling_2017}.

\subsection{Data Aggregation for Time Series Visualization}
Data aggregation for time series visualization can be categorized into density-wise data aggregation and downsampling.
Density-wise data aggregation employs a shared color-coding to generate an image of the data on the server-side, which is transmitted to and displayed on the client front-end~\cite{datashader}. 
Downsampling reduces the number of data points that are transmitted to the visualization front-end while aiming to preserve specific characteristics or the overall shape of the data. Consequently, this approach results in reduced latency, enhanced client-side rendering time, and increased responsiveness, as the application is not burdened with rendering all data points~\cite{agrawal_challenges_2015}.

Downsampling can be further differentiated into characteristic and value preserving data aggregation. 
Characteristic data aggregation aims to emphasize specific properties or trends in the data by employing aggregation operations such as mean, median, or smoothing~\cite{hellerstein1999control, rong_asap_2017}. 
Conversely, value preserving data aggregation, also referred to as data point selection, selects data points from the original time series with the objective of preserving its overall shape. 

\subsubsection{Value Preserving Data Aggregation}
Numerous value preserving data aggregation algorithms have been proposed in literature~\cite{bae2017practical, gil2021towards}. Among these, EveryNth, MinMax, M4~\cite{jugel_m4_2014}, and LTTB~\cite{steinarsson_downsampling_nodate} are arguably the most prevalent algorithms~\cite{van2023data, timescaleblog_paganini_2023, uber_raskin_aggarwal_2018}. 

Table~\ref{tab:aggregators_overview_} presents an overview of the computational properties of these four algorithms.
The EveryNth algorithm selects every $n^{th}$ data point in order to construct an output of $n_{out}$ data points, which is an inexpensive and common query operation~\cite{jugel_vdda_2016}. 
MinMax downsampling involves selecting the vertical extrema for each bin (i.e., bucket $B_i$), using the (arg)min and (arg)max operations, which are also highly optimized queries~\cite{tangwongsan2015general}. 
M4 can be viewed as a combination of EveryNth and MinMax, essentially selecting the vertical (min and max) and horizontal (first and last) extrema for each bucket~\cite{jugel_m4_2014}. 
Given the highly optimized server-side query operations that these three algorithms capitalize on and the relatively low cost of the operations involved (selecting or comparing data), improving the computational properties of these three algorithms is practically infeasible.
It is important to note that in-memory approaches also benefit from the low computational cost and the parallelizability of these algorithms.

LTTB is based on the concept of effective triangular areas, which is often employed in line simplification algorithms~\cite{steinarsson_downsampling_nodate}. Specifically, LTTB selects in each bucket $B_i$ the data point that forms the largest triangular surface with the previously selected data point and the next bucket’s ($B_{i+1}$) average value.
In contrast to the above three algorithms, LTTB involves more expensive operations, including (i) computing the average for each bin, and (ii) calculating and comparing the triangular surface for each data point within a bin. Furthermore, this algorithm requires a sequential pass over the data, making parallelization impossible. Implementing LTTB in database solutions also requires a user-defined function, benefiting less from server-side query optimizations~\cite{gjengset2018noria}. As such, LTTB is considerably more expensive in both out-of-core (i.e., query) and in-memory contexts.

Given these observations on LTTB's computational properties, the focus of this work is to improve the scalability of LTTB, striving towards the same computational properties as M4 and MinMax.


\begin{table}[]
\centering
\caption{Overview of time series data point selection algorithms, where $N$ denotes the time series length, and $n_{out}$ represents the aggregation output size. We included the MinMaxLTTB algorithm for comparison.}
\label{tab:aggregators_overview_}
\resizebox{\columnwidth}{!}{%
    \begin{tabular}{@{}rcccc@{}}
    \toprule
    \textbf{} & \textbf{\begin{tabular}[c]{@{}c@{}}(i) \\ Time\end{tabular}} & \textbf{\begin{tabular}[c]{@{}c@{}}(ii) \\ Memory\end{tabular}} & \textbf{\begin{tabular}[c]{@{}c@{}}(iii) \\ Parallelizable\end{tabular}}   &
    \textbf{\begin{tabular}[c]{@{}c@{}}(iv) \\ Output Memory\end{tabular}} \\ \midrule
    EveryNth & $O(n_{out})$ & $O(1)$ & \checkmark & $O(n_{out})$ \\
    MinMax & $O(N)$ & $O(1)$ & \checkmark & $O(n_{out})$ \\
    M4 \cite{jugel_m4_2014} & $O(N)$ & $O(1)$ & \checkmark & $O(n_{out})$ \\
    LTTB \cite{steinarsson_downsampling_nodate} & $O(N)$ & $O(1)$ & $\times$ & $O(n_{out})$ \\
    \midrule
    MinMaxLTTB & $O(N)$ & $O(n_{out})$ & \checkmark & $O(n_{out})$ \\
    \bottomrule
    \vspace{-5mm}
    \end{tabular}
}
\end{table}

\section{MinMaxLTTB}
\begin{figure*}[!tb]
 \centering 
 \includegraphics[width=.995\linewidth]{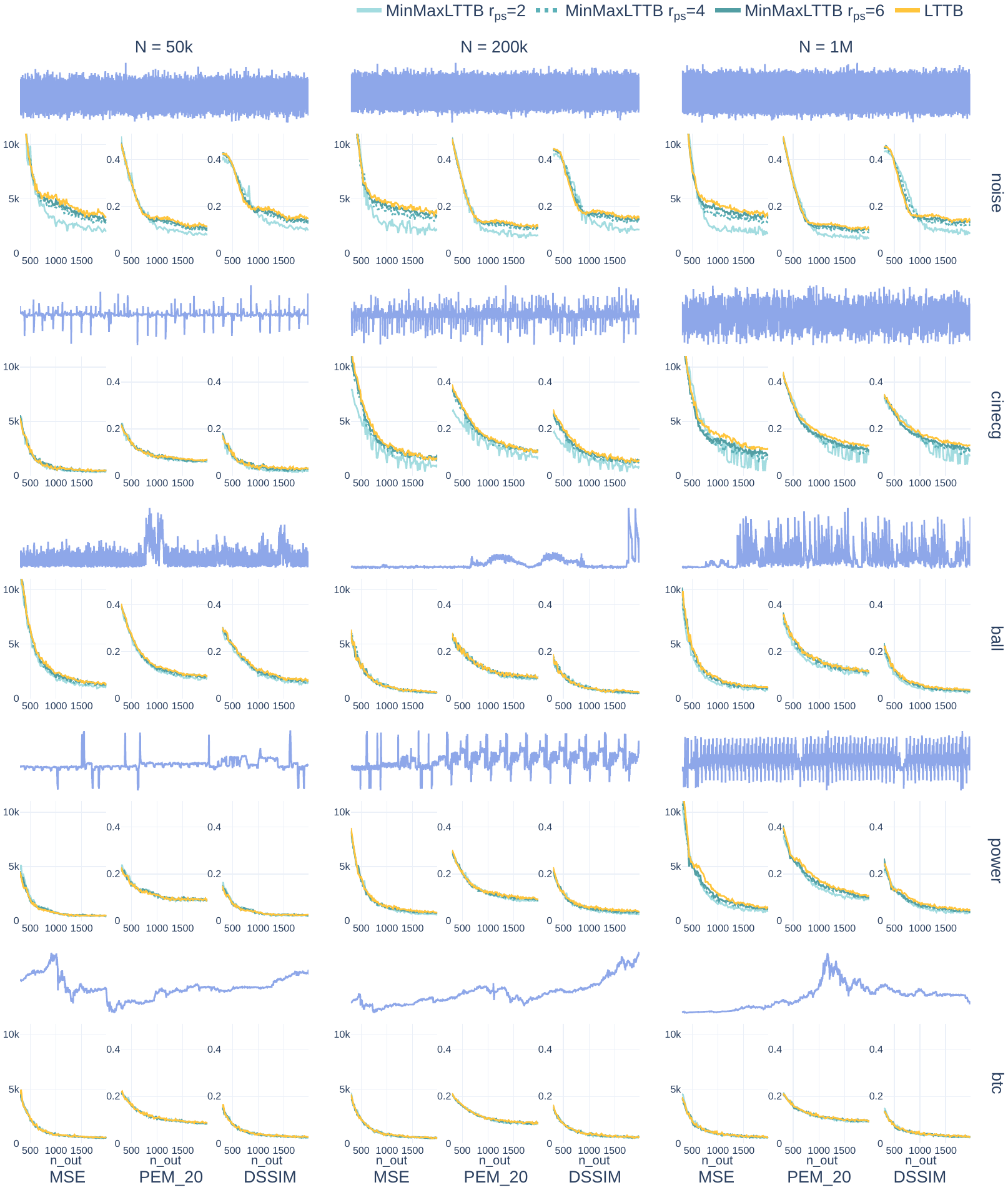}
 \vspace*{-3mm}
 \caption{Assessing visual representativeness of \texttt{MinMaxLTTB} with $r_{ps} \in \{2, 4, 6\}$ for various time series templates. Each row displays a distinct time series dataset, with columns indicating the template size. All image templates, including the blue reference templates which are depicted above the metric subplots, were generated using Plotly’s default settings (linear interpolation, line-width of 2 pixels). The metric subplots reveal trends in aggregation algorithm performance as $n_{out}$ (x-axis) increases (range [200, 2000]). More information about these trends, metrics, conv-mask scaling, and  templates can be found in~\cite{van2023data}. 
 $PEM\_20$ refers to Pixel Error with a Margin of 20, where pixels differences above 20 are binarized and divided by the conv-mask size to obtain a ratio. DSSIM denotes conv-mask scaled structural dissimilarity, while MSE represents conv-mask scaled Mean Squared Error. A \href{https://github.com/predict-idlab/MinMaxLTTB/blob/main/gifs/visual_representativeness_preselection_ratio.gif}{GIF} and an \href{https://github.com/predict-idlab/MinMaxLTTB/blob/main/animations/frame_preselection_ratio.html}{HTML animation} further demonstrate the visual representativeness of MinMaxLTTB for $r_{ps} \in [1-8]$.
}
 \vspace*{-3mm}
 \label{fig:vis_repr}
\end{figure*}

\begin{figure}[!tb]
 \centering 
 \includegraphics[width=\linewidth]{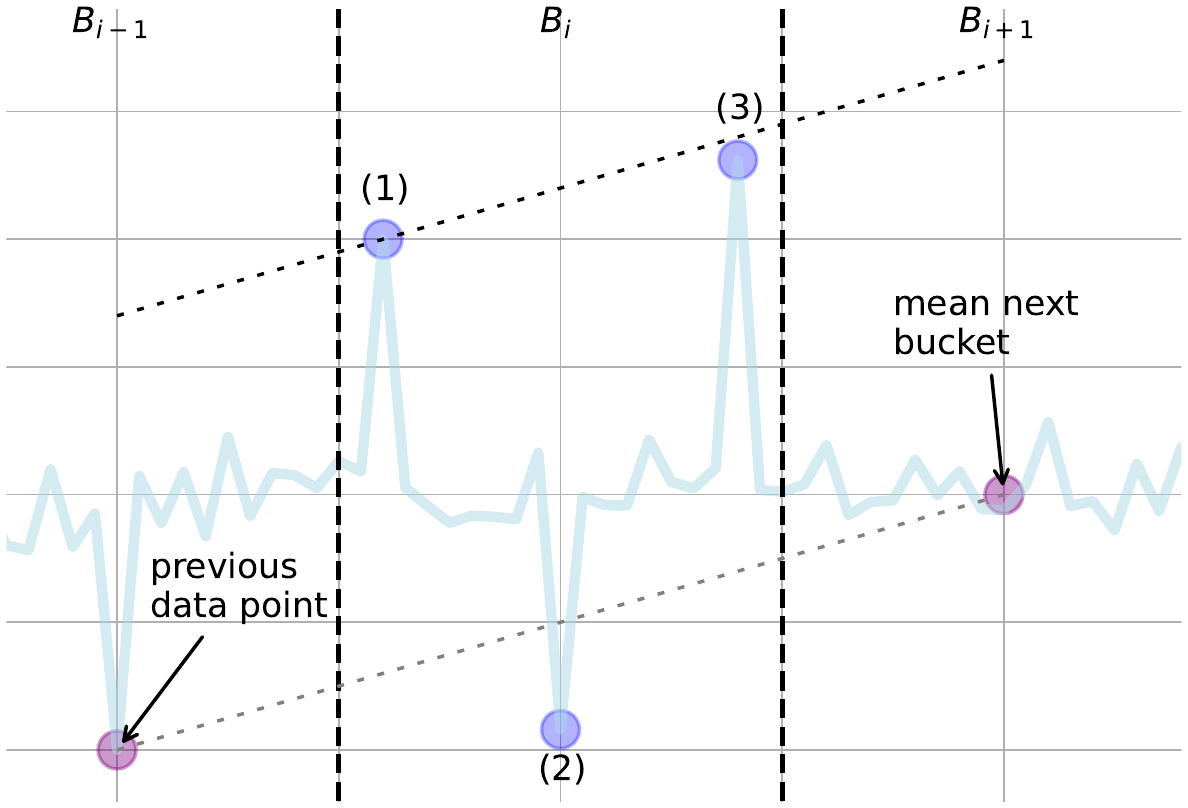}
 \vspace*{-5mm}
 \caption{
  Illustration of LTTB's tendency to (i) select contrasting extrema values in neighboring buckets and  
  (ii) favor extrema proximity to the left-bin edge.
  For all points in $B_i$, triangular areas are calculated using the prior selected point from $B_{i-1}$ and the mean x and y values of $B_{i+1}$. 
  Points (1) and (3), both opposing the selected extrema of $B_{i-1}$, compete for the largest triangular surface. Note that extremum (2) would yield a considerably smaller triangle than points (1) and (3). The dashed line passing through point (1) represents the \textit{equisurface} line, with points above it generating larger triangles and those below resulting in smaller ones. This line will always be parallel to the one connecting the previous data point and the mean of the next bucket. Despite being the global extrema of $B_i$, (3) is not selected as it resides below the equisurface line of (1), illustrating LTTB's tendency to favor data points near the left-bin edge.
 }
 \label{fig:lttb_surface}
\end{figure}


We propose MinMaxLTTB, an enhancement to LTTB that addresses its unfavorable computational properties. This is achieved by building on insights from prior research, which indicated the importance of selecting (alternating) vertical extrema for representative data aggregation~\cite{van2023data, steinarsson_downsampling_nodate, jugel_m4_2014}. 
MinMaxLTTB is realized in a two-step process, where we (i) \textit{preselect} vertical extrema by using MinMax, and then (ii) \textit{apply} LTTB on the preselected data points. 
In addition to the MinMax-preselection, the first and last data point of the original time series are also passed to the LTTB algorithm.

In the first step, we preselect $r_{ps} \cdot n_{out}$ data points, where $r_{ps} \geq 2$ denotes the integer preselection ratio and $n_{out}$ the number of output data points. Remark that $r_{ps} / 2$ indicates the number of LTTB \textit{sub-buckets} for which vertical extrema (i.e., min and max) will be selected, e.g., an $r_{ps}$ of 8 can be interpreted as dividing each LTTB bucket into $4$ sub-buckets~\footnote{Note that MinMaxLTTB with $r_{ps} = 1$ corresponds to MinMax aggregation.}. 

In the second step, the LTTB algorithm is applied on the $r_{ps} \cdot n_{out}$ data points, allowing LTTB to scale with $n_{out}$ instead of the time series length $N$. 
Although the MinMax algorithm scales with $N$, it is relatively inexpensive and easily parallelizable, facilitating efficient data reduction. Furthermore, since extracting the (arg)min and (arg)max is a common query, this technique can be seamlessly integrated into database solutions where it benefits from server-side query optimizations such as caching~\cite{gjengset2018noria}.


%


\subsection{Visual Representativeness and Preselection Ratio}

We analyze the visual representativeness of MinMaxLTTB for various preselection ratios $r_{ps} \in \{2, 4, 6 \}$ using the methodology proposed in~\cite{van2023data}. We refer the reader to this prior work for details on the visual representativeness metrics and the selected time series templates.

A first key observation is that MinMaxLTTB's performance curves are on par with LTTB for low-roughness series such as the btc templates, power $\leq$200k, cinecg-50k, and ball-200k. In these cases, the MinMax-preselection ratio has little influence, possibly explained by the absence of prominent extrema in the templates.
For higher roughness series such as the noise templates and cinecg $\geq$200k, MinMaxLTTB even seems to perform (slightly) better, especially for low $r_{ps} \in \{2, 4\}$.
This observation can be attributed to LTTB favoring the selection of data points near the left bin-edge. Figure~\ref{fig:lttb_surface} intuitively demonstrates that extrema near the left bin-edge allow creating larger triangular surfaces (with the selected extremum from the previous bucket), potentially omitting larger extrema that occur in the center or right part of the bin. Since MinMax-preselection with a low $r_{ps}$ preselects fewer options, while only considering the vertical position, LTTB's tendency to select near left-bin edge extrema values is partly mitigated.
For example, in Figure~\ref{fig:lttb_surface}, using $r_{ps} = 2$, points (2) and (3) would get preselected for bucket $B_i$, resulting in MinMaxLTTB selecting the bin-maximum (3) instead of (1). This results in improved visual representativeness (compared to the reference template) as more prominent data points are retained.

Consequently, increasing the preselection ratio makes the MinMaxLTTB performance curves less noisy and shift more towards the LTTB curves, as MinMaxLTTB will have more near left bin-edge preselected extrema. In summary, MinMaxLTTB does not degrade in visual representativeness with respect to visual LTTB, and a low $r_{ps} \in \{4, 6\}$ already results in high visual similarity to LTTB.


\subsection{Performance}

\begin{figure}[!tb]
    \centering
    \includegraphics[width=\linewidth]{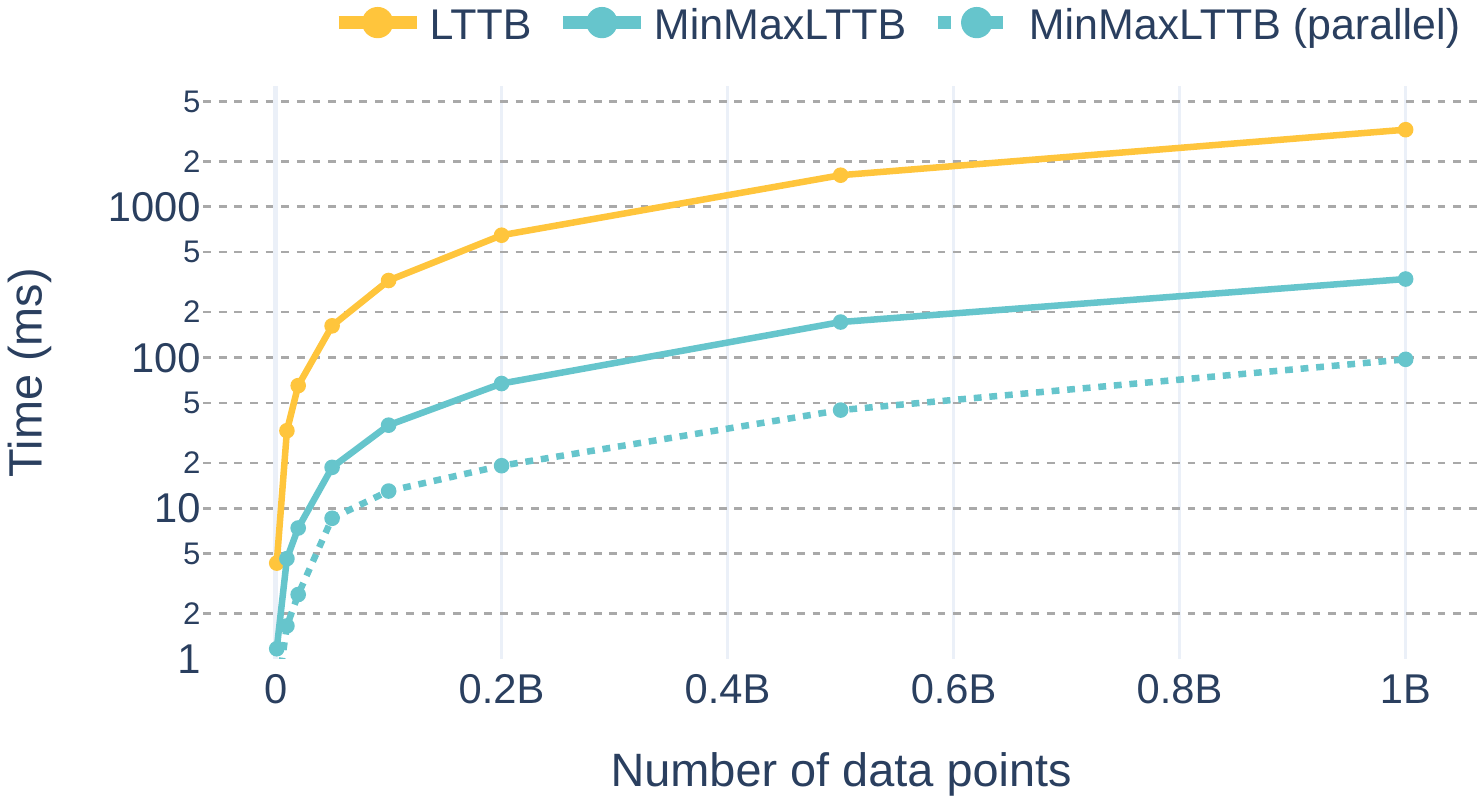}
    \caption{In-memory performance analysis of LTTB and MinMaxLTTB. For LTTB the \href{https://github.com/predict-idlab/plotly-resampler/blob/c60833eb4382b3072cb591b36de737dbf769a5ea/plotly_resampler/aggregation/algorithms/lttb_c.py}{C implementation from \texttt{plotly-resampler} v0.8.3.2} is used, for MinMaxLTTB the implementation can be found \href{https://github.com/predict-idlab/tsdownsample}{here}.}
    \label{fig:performance}
    \vspace{-4mm}
\end{figure}




MinMaxLTTB exhibits the same linear time complexity as LTTB (see Table~\ref{tab:aggregators_overview_}). The memory complexity of MinMaxLTTB is $O(n_{out})$  instead of $O(1)$ because, after the first step, $r_{ps} \cdot n_{out}$ indices are preselected (and thus stored in memory). 
In the previous section, it was demonstrated that a low $r_{ps}$ is sufficient to achieve comparable visual representativeness. Therefore, the additional memory overhead associated with MinMaxLTTB is almost negligible compared to LTTB, since the memory complexity of constructing the output is $O(n_{out})$~\footnote{Note that this cost is unavoidable for all algorithms, and therefore included in the output memory column of Table~\ref{tab:aggregators_overview_}.}.

Although the runtime of both algorithms scales linearly with the time series size ($N$), the slope of this linear scaling is significantly lower (i.e., better) for MinMaxLTTB. This can intuitively be explained by looking at the operations that are performed (for each data point) by both algorithms. In particular, both algorithms perform comparisons, with LTTB comparing for the largest triangular surface and MinMax comparing for extrema. However, in addition to a comparison, LTTB's operations include the calculation of the triangular surfaces (and the bin-wise average), thus requiring more operations than MinMax,
which results in a higher slope for LTTB.
In addition to superior linear scaling, MinMaxLTTB also allows for parallelization, which further alleviates the linear scaling to multiple cores. 
Figure~\ref{fig:performance} demonstrates this improved scaling of MinMaxLTTB, enhancing the performance with an order of magnitude (10x). Furthermore, when applying parallelization on MinMaxLTTB, the performance increases 30x compared to LTTB.



\section{Conclusion}
This work proposes MinMaxLTTB, a downsampling technique that mitigates the unfavorable computational properties of LTTB through the use of MinMax-preselection. Our evaluation of MinMaxLTTB's visual representativeness reveals that even a small preselection ratio $r_{ps} \in \{4, 6\}$ yields a high degree of similarity to LTTB. Performance analysis demonstrated that MinMaxLTTB's computation time decreases by over an order of magnitude compared to LTTB. This is particularly significant for scalable visualization of big data, as interactive latency greatly influences the rate at which users make observations during exploratory analysis~\cite{liu2014effects}. 
We further hypothesize that the demonstrated effectiveness of MinMax-preselection could potentially be extended to other computationally demanding data point selection algorithms such as Visvalingam-Whyat~\cite{visvalingam1993line}, Douglas-Peucker~\cite{douglas1973algorithms}, and Longest-Line-Bucket~\cite{steinarsson_downsampling_nodate}.
Finally, the fact that MinMaxLTTB is currently the default aggregation algorithm for already over half a year in a widely utilized visualization tool serves as strong evidence of its efficacy.




\acknowledgments{%
    The authors wish to thank Louise Van Calenbergh for proofreading the manuscript.
    Jonas Van Der Donckt (1S56322N) is funded by a doctoral fellowship of the Research Foundation Flanders (FWO).
    Part of this work is done in the scope of the imec.AAA Context-aware health monitoring project. 
}

\bibliographystyle{abbrv-doi}

\bibliography{template}

\begin{thebibliography}{10}

\bibitem{agrawal_challenges_2015}
R.~Agrawal, A.~Kadadi, X.~Dai, and F.~Andres.
\newblock Challenges and opportunities with big data visualization.
\newblock In {\em Proceedings of the 7th {International} {Conference} on
  {Management} of computational and collective {intElligence} in {Digital}
  {EcoSystems}}, pp. 169--173. ACM, Caraguatatuba Brazil, Oct. 2015. doi: {{%
10\hspace{.1pt}\discretionary{.}{%
}{.}\hspace{.4pt}1145\discretionary{/}{%
}{/}2857218\hspace{.1pt}\discretionary{.}{%
}{.}\hspace{.4pt}2857256}}


\bibitem{aigner_visualizing_2007}
W.~Aigner, S.~Miksch, W.~Müller, H.~Schumann, and C.~Tominski.
\newblock Visualizing time-oriented data—{A} systematic view.
\newblock {\em Computers \& Graphics}, 31(3):401--409, June 2007.
\newblock ZSCC: 0000450. doi: {{%
10\hspace{.1pt}\discretionary{.}{%
}{.}\hspace{.4pt}1016\discretionary{/}{%
}{/}j\hspace{.1pt}\discretionary{.}{%
}{.}\hspace{.4pt}cag\hspace{.1pt}\discretionary{.}{%
}{.}\hspace{.4pt}2007\hspace{.1pt}\discretionary{.}{%
}{.}\hspace{.4pt}01\hspace{.1pt}\discretionary{.}{%
}{.}\hspace{.4pt}030}}


\bibitem{bae2017practical}
P.~Bae, K.-W. Lim, W.-S. Jung, and Y.-B. Ko.
\newblock Practical implementation of m4 for web visualization service.
\newblock {\em Journal of Communications and Networks}, 19(4):384--391, 2017.
  doi: {{%
10\hspace{.1pt}\discretionary{.}{%
}{.}\hspace{.4pt}1109\discretionary{/}{%
}{/}JCN\hspace{.1pt}\discretionary{.}{%
}{.}\hspace{.4pt}2017\hspace{.1pt}\discretionary{.}{%
}{.}\hspace{.4pt}000062}}


\bibitem{bikakis_big_2018}
N.~Bikakis.
\newblock Big {Data} {Visualization} {Tools}.
\newblock {\em arXiv:1801.08336 [cs]}, Feb. 2018.
\newblock ZSCC: 0000075 arXiv: 1801.08336.

\bibitem{caldarola2017big}
E.~G. Caldarola and A.~M. Rinaldi.
\newblock Big data visualization tools: a survey.
\newblock {\em Research Gate}, 2017.

\bibitem{timescaleblog_paganini_2023}
J.~Davi~Paganini.
\newblock Downsampling in the database: How data locality can improve data
  analysis.
\newblock
  \href{https://www.timescale.com/blog/downsampling-in-the-database-how-data-locality-can-improve-data-analysis/}{\nolinkurl{www.timescale.com/blog}},
  Feb 2023.

\bibitem{douglas1973algorithms}
D.~H. Douglas and T.~K. Peucker.
\newblock Algorithms for the reduction of the number of points required to
  represent a digitized line or its caricature.
\newblock {\em Cartographica: the international journal for geographic
  information and geovisualization}, 10(2):112--122, 1973. doi: {{%
10\hspace{.1pt}\discretionary{.}{%
}{.}\hspace{.4pt}3138\discretionary{/}{%
}{/}FM57\discretionary{%
}{-}{-}6770\discretionary{%
}{-}{-}U75U\discretionary{%
}{-}{-}7727}}


\bibitem{gil2021towards}
A.~Gil, M.~Quartulli, I.~G. Olaizola, and B.~Sierra.
\newblock Towards smart data selection from time series using statistical
  methods.
\newblock {\em IEEE Access}, 9:44390--44401, 2021. doi: {{%
10\hspace{.1pt}\discretionary{.}{%
}{.}\hspace{.4pt}1109\discretionary{/}{%
}{/}ACCESS\hspace{.1pt}\discretionary{.}{%
}{.}\hspace{.4pt}2021\hspace{.1pt}\discretionary{.}{%
}{.}\hspace{.4pt}3066686}}


\bibitem{gjengset2018noria}
J.~Gjengset, M.~Schwarzkopf, J.~Behrens, L.~T. Ara{\'u}jo, M.~Ek, E.~Kohler,
  M.~F. Kaashoek, and R.~T. Morris.
\newblock Noria: dynamic, partially-stateful data-flow for high-performance web
  applications.
\newblock In {\em OSDI}, vol.~18, pp. 213--231, 2018.

\bibitem{godfrey2016interactive}
P.~Godfrey, J.~Gryz, P.~Lasek, and N.~Razavi.
\newblock Interactive visualization of big data.
\newblock In {\em Beyond Databases, Architectures and Structures. Advanced
  Technologies for Data Mining and Knowledge Discovery: 12th International
  Conference, BDAS 2016, Ustro{\'n}, Poland, May 31-June 3, 2016, Proceedings
  11}, pp. 3--22. Springer, 2016.

\bibitem{hellerstein1999control}
J.~M. Hellerstein, R.~Avnur, A.~Chou, C.~Hidber, C.~Olston, V.~Raman, T.~Roth,
  and P.~J. Haas.
\newblock Interactive data analysis: The control project.
\newblock {\em Computer}, 32(8):51--59, 1999. doi: {{%
10\hspace{.1pt}\discretionary{.}{%
}{.}\hspace{.4pt}1109\discretionary{/}{%
}{/}2\hspace{.1pt}\discretionary{.}{%
}{.}\hspace{.4pt}781635}}


\bibitem{datashader}
{Holoviz}-community.
\newblock Datashader, quickly and accurately render even the largest data.
\newblock \url{https://github.com/holoviz/datashader}.

\bibitem{jugel_m4_2014}
U.~Jugel, Z.~Jerzak, G.~Hackenbroich, and V.~Markl.
\newblock M4: a visualization-oriented time series data aggregation.
\newblock {\em Proceedings of the VLDB Endowment}, 7(10):797--808, June 2014.
  doi: {{%
10\hspace{.1pt}\discretionary{.}{%
}{.}\hspace{.4pt}14778\discretionary{/}{%
}{/}2732951\hspace{.1pt}\discretionary{.}{%
}{.}\hspace{.4pt}2732953}}


\bibitem{jugel_vdda_2016}
U.~Jugel, Z.~Jerzak, G.~Hackenbroich, and V.~Markl.
\newblock {VDDA}: automatic visualization-driven data aggregation in relational
  databases.
\newblock {\em The VLDB Journal}, 25:53--77, Feb. 2016. doi: {{%
10\hspace{.1pt}\discretionary{.}{%
}{.}\hspace{.4pt}1007\discretionary{/}{%
}{/}s00778\discretionary{%
}{-}{-}015\discretionary{%
}{-}{-}0396\discretionary{%
}{-}{-}z}}


\bibitem{kumar_mr_2021}
S.~Kumar, M.~P. Andersen, and D.~E. Culler.
\newblock Mr. {Plotter}: {Unifying} {Data} {Reduction} {Techniques} in
  {Storage} and {Visualization} {Systems}, June 2021.
\newblock arXiv:2106.12505 [cs].

\bibitem{kwon_sampling_2017}
B.~C. Kwon, J.~Verma, P.~J. Haas, and C.~Demiralp.
\newblock Sampling for {Scalable} {Visual} {Analytics}.
\newblock {\em IEEE Computer Graphics and Applications}, 37(1):100--108, Jan.
  2017. doi: {{%
10\hspace{.1pt}\discretionary{.}{%
}{.}\hspace{.4pt}1109\discretionary{/}{%
}{/}MCG\hspace{.1pt}\discretionary{.}{%
}{.}\hspace{.4pt}2017\hspace{.1pt}\discretionary{.}{%
}{.}\hspace{.4pt}6}}


\bibitem{lin_visualizing_2005}
J.~Lin, E.~Keogh, and S.~Lonardi.
\newblock Visualizing and {Discovering} {Non}-{Trivial} {Patterns} in {Large}
  {Time} {Series} {Databases}.
\newblock {\em Information Visualization}, 4(2):61--82, June 2005.
\newblock ZSCC: 0000178. doi: {{%
10\hspace{.1pt}\discretionary{.}{%
}{.}\hspace{.4pt}1057\discretionary{/}{%
}{/}palgrave\hspace{.1pt}\discretionary{.}{%
}{.}\hspace{.4pt}ivs\hspace{.1pt}\discretionary{.}{%
}{.}\hspace{.4pt}9500089}}


\bibitem{liu2014effects}
Z.~Liu and J.~Heer.
\newblock The effects of interactive latency on exploratory visual analysis.
\newblock {\em IEEE transactions on visualization and computer graphics},
  20(12):2122--2131, 2014.

\bibitem{rostislav_performance_javascript_vis}
R.~Netek, J.~Brus, and O.~Tomecka.
\newblock Performance testing on marker clustering and heatmap visualization
  techniques: A comparative study on javascript mapping libraries.
\newblock {\em ISPRS International Journal of Geo-Information}, 8(8), 2019.
  doi: {{%
10\hspace{.1pt}\discretionary{.}{%
}{.}\hspace{.4pt}3390\discretionary{/}{%
}{/}ijgi8080348}}


\bibitem{uber_raskin_aggarwal_2018}
B.~Raskin and N.~Aggarwal.
\newblock The billion data point challenge: Building a query engine for high
  cardinality time series data.
\newblock
  \href{https://www.uber.com/en-BE/blog/billion-data-point-challenge/}{\nolinkurl{uber.com/billion-data-point-challenge}},
  Dec 2018.

\bibitem{rong_asap_2017}
K.~Rong and P.~Bailis.
\newblock {ASAP}: prioritizing attention via time series smoothing.
\newblock {\em Proceedings of the VLDB Endowment}, 10(11):1358--1369, Aug.
  2017. doi: {{%
10\hspace{.1pt}\discretionary{.}{%
}{.}\hspace{.4pt}14778\discretionary{/}{%
}{/}3137628\hspace{.1pt}\discretionary{.}{%
}{.}\hspace{.4pt}3137645}}


\bibitem{shneiderman_eyes_nodate}
B.~Shneiderman.
\newblock The {Eyes} {Have} {It}: {A} {Task} by {Data} {Type} {Taxonomy} for
  {Information} {Visualizations}.
\newblock In {\em Proceedings 1996 IEEE symposium on visual languages}, pp.
  336--343. IEEE, 1996. doi: {{%
10\hspace{.1pt}\discretionary{.}{%
}{.}\hspace{.4pt}1016\discretionary{/}{%
}{/}B978\discretionary{%
}{-}{-}155860915\discretionary{%
}{-}{-}0\discretionary{/}{%
}{/}50046\discretionary{%
}{-}{-}9}}


\bibitem{steinarsson_downsampling_nodate}
S.~Steinarsson.
\newblock Downsampling {Time} {Series} for {Visual} {Representation}.
\newblock Master's thesis, University of Iceland, 2013. doi: {{%
1946\discretionary{/}{%
}{/}15343}}


\bibitem{tangwongsan2015general}
K.~Tangwongsan, M.~Hirzel, S.~Schneider, and K.-L. Wu.
\newblock General incremental sliding-window aggregation.
\newblock {\em Proceedings of the VLDB Endowment}, 8(7):702--713, 2015.

\bibitem{van2022plotly}
J.~Van Der~Donckt, J.~Van Der~Donckt, E.~Deprost, and S.~Van~Hoecke.
\newblock Plotly-resampler: Effective visual analytics for large time series.
\newblock In {\em 2022 IEEE Visualization and Visual Analytics (VIS)}, pp.
  21--25. IEEE, 2022. doi: {{%
10\hspace{.1pt}\discretionary{.}{%
}{.}\hspace{.4pt}1109\discretionary{/}{%
}{/}VIS54862\hspace{.1pt}\discretionary{.}{%
}{.}\hspace{.4pt}2022\hspace{.1pt}\discretionary{.}{%
}{.}\hspace{.4pt}00013}}


\bibitem{van2023data}
J.~Van Der~Donckt, J.~Van Der~Donckt, M.~Rademaker, and S.~Van~Hoecke.
\newblock Data point selection for line chart visualization: Methodological
  assessment and evidence-based guidelines.
\newblock {\em arXiv preprint arXiv:2304.00900}, 2023.

\bibitem{visvalingam1993line}
M.~Visvalingam and J.~D. Whyatt.
\newblock Line generalisation by repeated elimination of points.
\newblock {\em The cartographic journal}, 30(1):46--51, 1993. doi: {{%
10\hspace{.1pt}\discretionary{.}{%
}{.}\hspace{.4pt}1179\discretionary{/}{%
}{/}000870493786962263}}


\bibitem{walker_timenotes_2016}
J.~Walker, R.~Borgo, and M.~W. Jones.
\newblock {TimeNotes}: {A} {Study} on {Effective} {Chart} {Visualization} and
  {Interaction} {Techniques} for {Time}-{Series} {Data}.
\newblock {\em IEEE Transactions on Visualization and Computer Graphics},
  22(1):549--558, Jan. 2016. doi: {{%
10\hspace{.1pt}\discretionary{.}{%
}{.}\hspace{.4pt}1109\discretionary{/}{%
}{/}TVCG\hspace{.1pt}\discretionary{.}{%
}{.}\hspace{.4pt}2015\hspace{.1pt}\discretionary{.}{%
}{.}\hspace{.4pt}2467751}}


\end{thebibliography}
\end{document}